# MoSe$_2$/WS$_2$ heterojunction photodiode integrated with a silicon nitride waveguide for visible light detection with high responsivity

Rivka Gherabli[1], S.R.K.C. Indukuri[1], Roy Zektzer[1], Christian Frydendahl[1], and Uriel Levy[1,*]

[1] Department of Applied Physics, The Faculty of Science, The Center for Nanoscience and Nanotechnology, The Hebrew University of Jerusalem, Jerusalem 91904, Israel.

[*] Corresponding author: ulevy@mail.huji.ac.il

**Abstract**

We demonstrate experimentally the realization and the characterization of a chip scale integrated photodetector for the visible and the near infrared spectral regime based on the integration of a MoSe$_2$/WS$_2$ heterojunction on top of a silicon nitride waveguide. This configuration achieves high responsivity of ~1 A/W at the wavelength of 780 nm (indicating an internal gain mechanism), while suppressing the dark current to the level of ~50 pA, much lower as compared to a reference sample of just MoSe$_2$ without WS$_2$. We have measured the power spectral density of the dark current to be as low as ~$1 \times 10^{-12}$ A/Hz$^{0.5}$, from which we extract the noise equivalent power (NEP) to be ~$1 \times 10^{-12}$ W/Hz$^{0.5}$. To demonstrate the usefulness of the device, we use it for the characterization of the transfer function of a microring resonator that is integrated on the same chip as the photodetector. The ability to integrate local photodetectors on chip and to operate such devices with high performance at the visible and the near-infrared regime is expected to play a critical role in future integrated devices in the field of optical communications, quantum photonics, biochemical sensing and more.

**Introduction**

Integrated photonics is a fast-developing topic of research, with many promising future applications in communications [1–3], metrology [4–6] and quantum computing [7–9]. The grand vision is to integrate all the relevant devices and to ultimately construct self-contained systems with their own light sources and photodetectors on a chip. Establishing such self-contained systems would solve the otherwise very laborious challenge of aligning and bonding macroscopic photonic fibers to nanophotonic chips [10,11] and make such systems easy to implement in any setup.

Integrating light sources or photodetectors on a chip is a grand challenge mostly due to material mismatch. Typically, the material needed to be used for light emission or absorption is very different from the waveguide material which is specifically chosen to be as transparent as possible for the wavelength of choice in order to reduce propagation loss. The integration of different materials is typically implemented by methods such as epitaxial growth [12,13] or direct wafer bonding of e.g. III-V semiconductors directly onto the chip [14–16]. The use of each of these approaches involves significant challenges, such as atomic lattice mismatch between the different materials resulting in strain and dislocations, or the high temperatures needed for the processes, which in general, are prohibitive for other fabrication steps [17]. While photodetectors can also be fabricated by evaporating absorbing materials such as metals [18,19] and amorphous silicon [20] or by using defects states in the waveguide material [21], such devices are generally limited in their performance.

A promising remedy for this grand challenge is the use of 2D materials. These materials are easier to integrate with different target substrates, as their polymer-based dry/wet transfer processes do not rely on any chemical reactions or the formation of atomic bonds. On top of this, their flatness and their strong inter-atomic bonds make them extremely flexible and durable, allowing them to conform to a large variety of substrate geometries. As a result, it is no surprise that in recent years there has been a multitude of research results on integrating 2D materials with different photonic systems [22–26].

While bulk semiconductors are a fundamental building block of integrated photonics, 2D semiconductors can have a similar role as they can absorb and emit photons and their optical properties can be tuned by electrostatic charge [27,28], light [29], physical stress/tension [30], and engineered defect states [31]. Transition metal dichalcogenides (TMDCs) are a family of 2D semiconductors where a unit cell of the structure consists of one transition metal atom, like molybdenum or tungsten, and two chalcogenide atoms, such as selenium or sulfur. The TMDCs become direct bandgap semiconductors when thinned down to a single monolayer, otherwise, they are indirect bandgap semiconductors. For this reason, they have been used to fabricate a variety of active photonic devices such as modulators [32], detectors [33,34], and especially light sources [35].

Up to now, the bulk of these works have been focused on integrated silicon photonics, operating at the telecom wavelengths [22–24,26], and was mostly motivated by the needs of telecommunication. However, there is a growing interest in the visible wavelength regime which can be supported by silicon-nitride-based integrated photonics for diverse applications in metrology, quantum computing, and biology to name a few [36]. Examples of this can be seen in the burgeoning field of nanoscale atom-light interactions. Here, hot vapors of atomic gasses, like rubidium or cesium are interfaced together with integrated photonic components, such as atomic cladded waveguides (ACWGs) [37–41], resonators [42,43], and surface plasmons [44–46], to create novel chip-scale atomic devices. Likewise, integrating biological samples with photonic circuits can enable new and interesting applications [47,48]. Detection of visible light is also promising for integrated quantum photonic devices. Color centers in silicon carbide [49] and diamond [50] emit visible photons, and there is a large effort to integrate these potential qubits with nanoscale devices on a chip.

In league with this growing effort, we hereby report an on-chip photodetector in the visible band based on a van der Waals heterostructure consisting of two TMDCs, $MoSe_2$ and $WS_2$ integrated with silicon nitride nanowaveguides. As the $MoSe_2$ and $WS_2$ flakes used are respectively p- and n-type semiconductors, our photodetector is essentially a heterostructure pn-junction diode. This results in a device with excellent performance metric, namely low dark current (~50 pA) and high peak responsivity of nearly 1 A/W. Such high performance metrics are needed if integrated photodetectors are to be fused with e.g. atomic vapors, where the expected signal intensities are very low due to the fast onset of optical saturation of the atomic vapors with increased probe light intensity [39]. Our device has a time response of about 20 MHz, likely limited by the specific device geometry studied here, as the carrier dynamics predict a much faster response [33]. Finally, to demonstrate the usefulness of our photodetector, we use it to characterize the spectral transfer function of a microring resonator that is fabricated on the same chip and is coupled to the same nanowaveguide. Such integration between photodetectors and photonic elements is critical for the full realization of on-chip integrated optical devices and systems.

**Results and discussion**

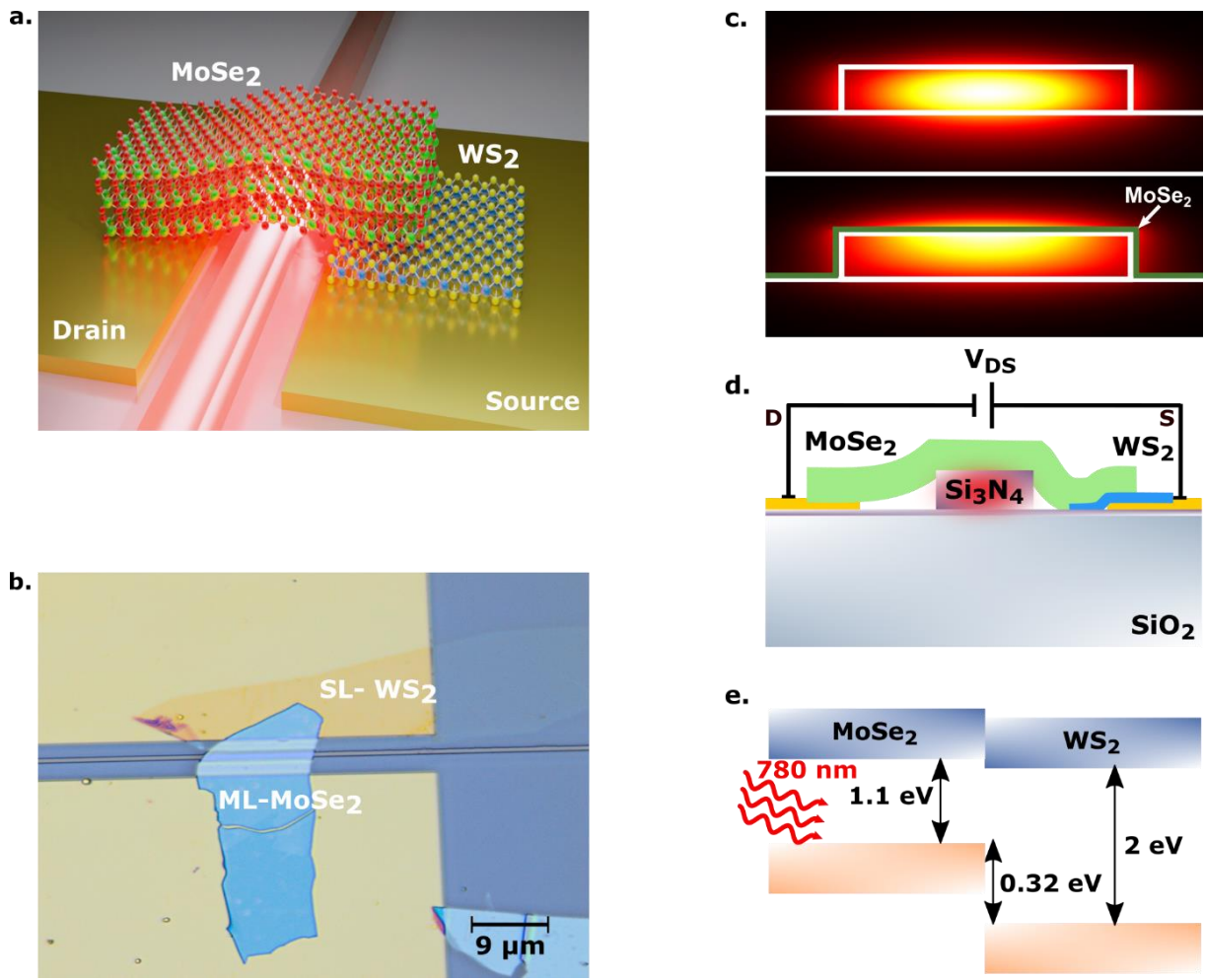

**Figure 1: a.** Schematic diagram of the MoSe$_2$-WS$_2$ heterostructure photodetector transferred on top of a silicon nitride waveguide. **b.** Optical microscope image of the fabricated photodetector. A single layer of WS$_2$ is transferred on top of one gold contact while the multilayer of MoSe$_2$ is placed on top of the silicon nitride waveguide **c.** Simulation of the electric field mode profile in the silicon nitride waveguide (top) and the silicon nitride waveguide with the MoSe$_2$ layers on top (bottom), (using Lumerical MODE software). The width of the silicon nitride waveguide is 700 nm. **d.** Schematic illustration of the device cross-section and the electrical configuration. **e.** Band diagram of bulk MoSe$_2$ and a single-layer WS$_2$, the two-layered materials form a type II heterojunction with a staggered gap.

A schematic of the device concept is shown in Fig. 1.a. The heterojunction photodetector consists of a bulk flake of molybdenum diselenide (MoSe$_2$) which is straddled over a silicon nitride waveguide implemented on the silicon chip. The MoSe$_2$ is brought into contact with a gold electrode (drain) on one side of the waveguide, and on the opposite side, it is in contact with a single layer (SL) flake of tungsten disulfide (WS$_2$), which is itself contacting a second gold electrode (source).

A microscope image of the actual device is shown in Fig. 1.b. As shown, the transferred single layer of WS$_2$ is in contact with one of the gold pads, whereas the MoSe$_2$ bulk flake of thickness of ~20 nm, is straddled on top of the second gold pad and the waveguide core region, covering ~13 μm length of the silicon nitride waveguide (SI Fig1). The ML-MoSe$_2$ extends to the SL-WS$_2$, without touching the source electrode. This way, the absorption of light propagating in the nanowaveguide is facilitated only by the bulk $MoSe_2$. 780 nm wavelength light is launched into the waveguide by using butt coupling via a lensed fiber. The light is then absorbed in the $MoSe_2$ multi-layers as the evanescent "tail" of the waveguide field mode spatially overlaps with the $MoSe_2$.

We simulated the fundamental mode of our waveguide structure (0.7 μm wide by 0.25 μm thick silicon nitride core on top of silicon oxide), with (photodetector waveguide mode) and without (unperturbed waveguide mode) the top layer of $MoSe_2$, as shown in Fig. 1.c. In the unperturbed waveguide, the electromagnetic field intensity resides mostly in the center of the waveguide core (Fig. 1.c top), while in the photodetector waveguide, the maximal electromagnetic field intensity is found to be at the interface between the $Si_3N_4$ and the $MoSe_2$ layers, (Fig 1.c bottom). This ensures high absorption in the $MoSe_2$ while maintaining a relatively high coupling efficiency (76%) between the unperturbed waveguide mode and the photodetector waveguide mode. From the simulation, we have extracted the propagation loss within the photodetector waveguide to be ~ $10^5$ dB/cm. This means that nearly 100% optical absorption (~20 dB) can be achieved with a device as short as ~ 2 μm.

Both TMDC layers were transferred as the last step of the device fabrication process. The flakes were transferred directly on top of the gold contacts in order to avoid contamination and doping of the 2D materials during the lithography step used for generating the contact pads. The TMDC heterojunction is made using a standard dry viscoelastic polymer transfer method [25], see methods below for more details on device fabrication in general. A sketch of the device cross-section structure can be seen in Fig. 1.d, where we also have defined the electrode configuration and the bias $V_{ds}$.

The $MoSe_2$ flake is p-doped while the $WS_2$ is n-doped. From observing the band diagram in Fig. 1.e we can see that the heterojunction is a PN type II (staggered gap) junction [51]. As the two conduction bands are more or less aligned, electrons see very little energy gap and can easily move between the flakes. On the other hand, the valence bands are misaligned by roughly 0.32 eV, effectively blocking the hole transfer from the $WS_2$ to the $MoSe_2$ side of the junction. This is indeed what should make the junction behave as a current rectifier, i.e. a diode. For this reason, we can operate the device in its reverse bias to minimize the measured dark current while still maximizing the collected photocurrent.

To validate the above mentioned expectations, we have measured the IV-characteristics of our device using a source measure unit (SMU), at dark and under illumination in the waveguide, see Fig. 2. For full details of the electrical and optical characterization, see methods. Indeed, we observe a clear region of forward (high dark current) and reverse (low dark current) bias, with a dark current as low as ~50 pA. This is in comparison to a reference sample, consisting of $MoSe_2$ without $WS_2$, where no rectification is observed, and the dark current is significantly higher (see SI Fig 4). Additionally, we see a large photocurrent that is generated when 780 nm light is introduced into the waveguide. Interestingly, we see that the rectification slowly breaks down for increasing optical powers (marked by red-colored curves in Fig. 2.a). The forward and reverse currents are nearly identical for 5 μW optical power, and for powers larger than this we even see that the forward and reverse bias actually flips. This effect is likely due to the asymmetric optical absorption in our device, as light is only absorbed in the $MoSe_2$ part of the junction (or may be due to optical absorption saturation). This is potentially causing a kind of gating effect altering the band alignment between the two materials in the junction.

Next, we plot the resulting photocurrent (total current minus the dark current) versus the optical power in the waveguide, for three representative reverse bias levels, see Fig. 2.b. We note that the behavior is sub-linear, indicating optical saturation effects in our device. If however, we restrict ourselves to the lowest powers used, we see a nearly linear regime of optical response with a responsivity of roughly 1 A/W for a reverse bias of 2 V, see Fig. 2.c. Such high responsivity is a clear indication of a gain mechanism in our device, as an ideal photodiode would have $R = 0.62$ A/W for a wavelength of 780 nm, i.e. $R = A\eta e/hf$ with absorption, $A$, and quantum efficiency, $\eta$, both equal 1.

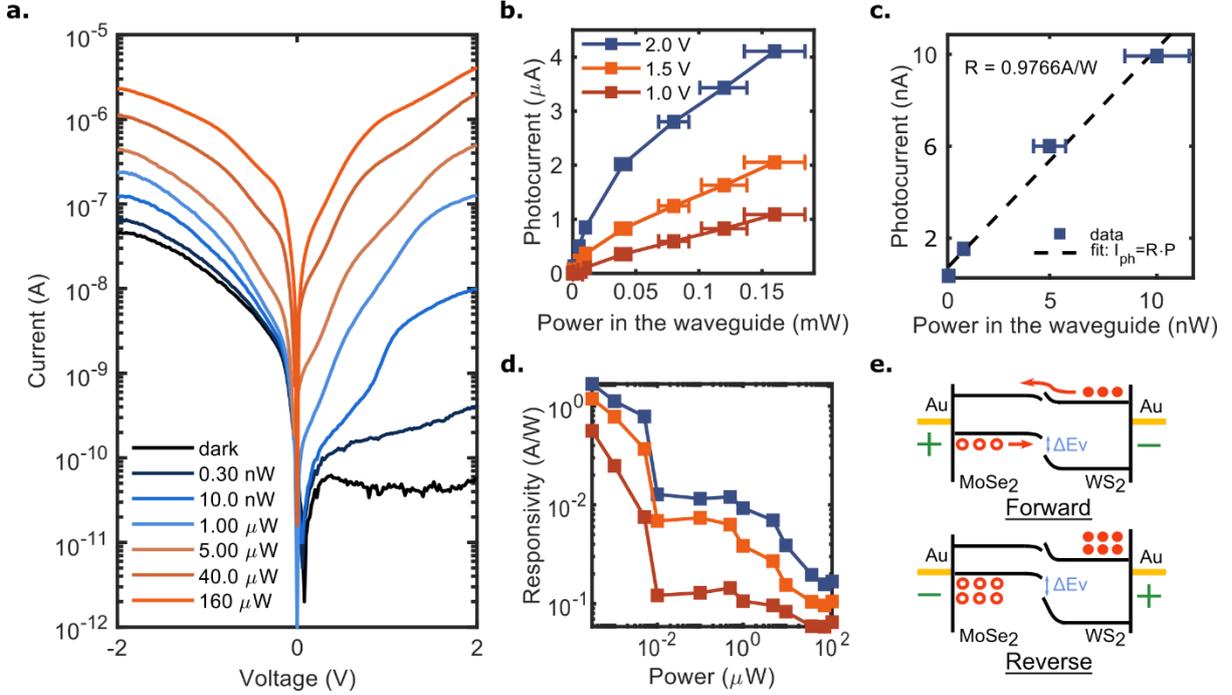

**Figure 2: a.** IV-characteristics of the diode for several optical powers in the waveguide. **b.** Photocurrent versus incident power on the diode through the waveguide. **c.** Responsivity for the low power (linear optical response) regime, showing near 1 A/W responsivity for a reverse bias of 2 V. **d.** Responsivity as a result of incident power, showing that the device saturates for even nominal powers. **e.** Band diagram of the MoSe$_2$-WS$_2$ heterojunction, showing charge transfer and the hole barrier at forward and reverse bias. The error bars for **b** and **c** are multiplied by a factor of 3.

Fig. 2.d shows the device's responsivity as a function of incident power. We see clearly that the responsivity decreases for larger optical powers, again indicating that optical saturation effects are non-negligible for our device for even nominal optical powers.

The heterojunction band structure diagram is shown in Fig. 2.e, at forward and reverse bias. The implementation of the photodetector with an asymmetrical contact was made possible through the introduction of a WS$_2$ monolayer between one gold contact and the bulk MoSe$_2$, where light absorption occurs. As explained earlier, with such a device configuration, there is a large band offset in the valence band of about $\Delta E_v = 0.32$ eV constituting a barrier for the hole carriers. This reduces the dark current without affecting the collection of photo-generated carriers (SI Fig 3). While the region of interest (the main area where absorption takes place) of the photodetector is the MoSe$_2$ multi-layer, the asymmetrical contact makes our device an excellent current rectifier (SI Fig 4). We have measured several devices and the results were consistent in all of them. Data for additional devices is given in the SI.

Next, we characterize the time/frequency response and noise characteristics of the device. Fig. 3.a shows the setup configuration for the time/frequency response measurements. In essence, the light coupled into the nanowaveguide via the lensed fiber is first amplitude/intensity modulated by a Mach-Zehnder Modulator (MZM). By altering the modulation frequency, we can observe the device's frequency response. For more details, see methods below. Fig. 3.b shows the device's frequency response. The modulation frequency of the incident light is gradually increased until we reach the point where we exceed the response time of the photodetector. From this measurement, we found the cutoff frequency (-3 dB point) to be ~20 MHz. The solid line shown in Fig. 3.b is a fit to a simple RC low-pass filter model. Likewise, when modulating the device's input light, we can measure the rise and fall times, see Fig. 3.c. The rise (defined by the time it takes the signal amplitude to increase from 10% to 90% of the maximum value), was found to be ~ 28 ns. Likewise, the fall time (defined by the time it takes the signal amplitude to decrease from 90% to 10% of the maximum value) was found to be ~ 20 ns.

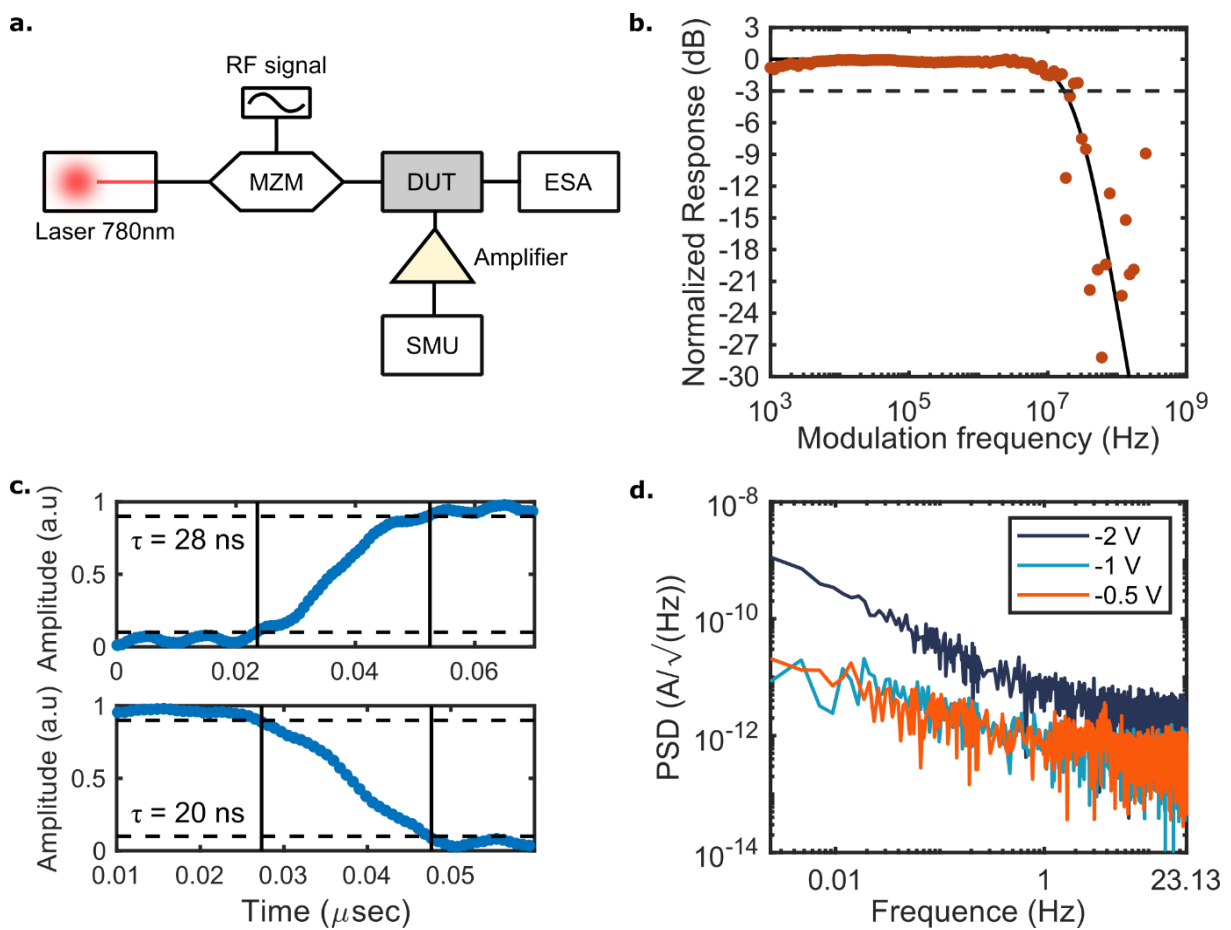

**Figure 3: a.** Measurement configuration for the device time response. The 780 nm laser that is coupled into the nanowaveguide is amplitude modulated with a Mach-Zehnder Modulator, and the modulation frequency response of the heterostructure detector is measured. **b.** Frequency response of the investigated photodetector. The modulation frequency of the light incident on the device is swept, and the device response -3 dB point is found at ~20 MHz. The solid line is a fit to a RC low-pass filter model. **c.** Rise and fall time measurements of the photodetector. We find a rise time of ~28 ns and a fall time of ~20 ns. **d.** Noise characteristics of the device. The PSD is measured for different bandwidth frequencies, and a PSD of ~1 ×10$^{-12}$ A/Hz$^{0.5}$ is found at a bandwidth of 1 Hz, corresponding to a device NEP of ~1 ×10$^{-12}$ W/Hz$^{0.5}$.

In order to analyze the noise characteristics of the device, we have measured the power spectral density (PSD) of the dark current. The device's dark current is monitored for 8 minutes, and the corresponding signal is Fourier transformed to get its frequency components (see Methods for full explanation). The results can be seen in Fig. 3.d. From the PSD we can find the Noise-Equivalent-Power (NEP) which is given by the PSD at 1 Hz divided by the device's responsivity, which is roughly ~1 A/W for low powers. We thus obtain a NEP of ~$1\times10^{-12}$ W/Hz$^{0.5}$. This NEP value is comparable to many of the commercial silicon based photodetectors.

As our photodetector is optimal for low power levels in the visible and the near infrared, showing good NEP with decent frequency response, and given its ease of integration on the chip, it is ideal for on-chip low-power applications such as monitoring optical components and operating in a low-power regime where saturation prohibits the use of high powers (e.g. in the case of atomic cladded waveguides). As such, we now demonstrate the usefulness of our on-chip device for the characterization of a microring resonator that is integrated on the same chip as the photodetector. Fig. 4a shows a schematic of how the microring resonator is coupled to the same waveguide as the photodetector device. Fig. 4b shows the detected photocurrent when the wavelength is scanned over a range of a few nanometers. Multiple resonances with a notch-like characteristic can easily be observed. Fig. 4c is a zoom in on a single resonance.

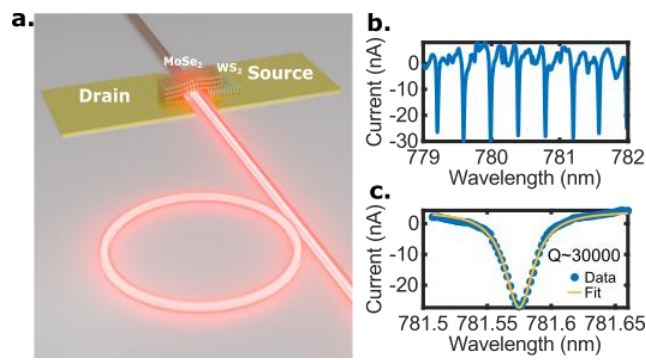

**Figure 4: a.** Schematic of an optical microring resonator coupled to the same nanowaveguide as the 2D material heterostructure photodetector. **b.** Measured photocurrent from the detector as the wavelength of the incident 780 nm laser is gradually detuned. **c.** Characterization of the MRR by fitting the measured photocurrent to the Lorentzian function.

From the measurements we can extract the free spectral range (FSR) of the 110 μm diameter microring resonator as 13.5 nm, which is in agreement with theory. We can also determine the full-width-half-maximum (FWHM) for a given resonance as to be 26 pm with a corresponding Q-factor of ~30000.

**Conclusions**

To conclude, we have demonstrated a chip scale photodetector with low dark current and high responsivity operating in the visible and the near infrared spectrum that is realized by integrating a heterostructure of the 2D material TMDCs MoSe$_2$ and WS$_2$ with a silicon nitride waveguide. The MoSe$_2$ consists of several layers, which ensures very high absorption of the incident radiation propagating along the silicon nitride waveguide. By adding a monolayer of WS$_2$ in between the MoSe$_2$ and the second gold electrode of the device, the dark current of the device is greatly suppressed when operating in the reverse bias, resulting in a heterostructure diode. We have measured a dark current as low as ~50 pA, significantly lower as compared to a pure MoSe$_2$ device without WS$_2$. When operating in the linear device response regime (i.e. at low intensities below optical saturation), we were able to achieve a responsivity as high as ~1 A/W for 780 nm light, indicating a slight gain mechanism in our device.

From frequency response measurements, we have found the 3 dB cutoff frequency to be ~20 MHz. The frequency response is limited most likely by the relatively large device geometry. We believe it is highly likely that a device with a frequency response in the GHz regime may be feasible using a slightly modified geometry. This line of thinking is based on the fact that the carrier dynamics of the materials predict much faster speeds [33].

By measuring the PSD of our device's dark current we find the NEP to be as low as ~1×10$^{-12}$ W/Hz$^{0.5}$. Taking advantage of this merit, we have used our photodetector to characterize the transfer function of an optical microring resonator that is integrated on the same chip as the photodetector. Being able to provide electrical measurements on a chip, we believe that our device demonstration is another step towards realizing fully on-chip integrated photonic devices. Additionally, the general performance metrics of the detector demonstrated here should also make it fully compatible with many of the growing topics in visible integrated photonics, such as waveguides integrated with atomic vapors, nano-diamond color center qubits, and biological lab-on-a-chip systems.

**Methods**

**Waveguide fabrication:** A 250 nm thick layer of silicon nitride is deposited by low-pressure chemical vapor deposition (LPCVD) on 2 μm thick thermal oxide. The waveguide cross-section is 0.7 µm width and 0.25 µm height. The waveguide geometry is defined by e-beam lithography and the structure is transferred from the resist to the silicon nitride by reactive ion etching [37]. Next, the wafers are diced such that it is possible to butt couple light into the waveguide's end facets. After this, another step of e-beam lithography is done to define the source and drain metal electrodes. The electrodes are made by evaporating 8 nm of Ti followed by 80 nm of Au using an evaporator and lift-off process.

**TMDC heterostructure fabrication:** Bulk flakes of MoSe$_2$ and WS$_2$ (HQ Graphene) are exfoliated via the Scotch tape method onto polydimethylsiloxane (PDMS) cut out from a Gel-Pak. Before exfoliation, the PDMS has been treated in an ozonator for 30 minutes to ensure minimal contamination from uncross-linked molecules from the PDMS. Next, the target flakes are found using an optical microscope and transferred to the intended waveguide by dry transfer technique with the help of home-built 2D transfer setup [52,53].

**Optical characterization:** To characterize the samples' electro-optic response, a 780 nm NewFocus TLB-67000 tunable laser source is butt coupled into the waveguide using a PM lensed fiber. Then, a Keysight B2901A SMU is connected to the gold electrodes using tungsten probes and 3D stages. Light at different power levels is injected into the waveguide, and the SMU performs a voltage sweep, $V_{ds}$, while measuring the current. To determine the laser power in the waveguide, the transmission through an identical waveguide without the 2D materials transferred on top is measured, collecting the transmitted light with another lensed fiber, and measuring the optical power by a Newport power meter.

**Time response measurements:** The time response measurements follow the same configuration as the other optical measurements above, with the exception that the input 780 nm laser is modulated by a iXblue Mach-Zehnder modulator (MZM). A sine RF signal is supplied to the modulator by a Keysight DSOX1004A digital storage oscilloscope, and the modulated signal is again butt coupled into the target waveguide. The modulated photoresponse from the heterostructure detector is then monitored on a Zurich Instruments digital lock-in amplifier oscilloscope.

## Author contributions

The project was conceived by all authors. Waveguide fabrication was done by R.Z. TMDC Heterostructures were fabricated by S.R.K.C.I. Optical and time response measurements were done by R.G., S.R.K.C.I. and R.Z. Data analysis was done by R.G. and C.F. U.L. supervised the project. All authors contributed to the writing of the manuscript and interpretation of the results. R.G., S.R.K.C.I contributed equally to the manuscript.

## Acknowledgements

R.G. acknowledges the support of the Shulamit Aloni scholarship of the Israeli Ministry of Science and Technology. C.F. is supported by the Carlsberg Foundation as an Internationalization Fellow. The research was partially supported by the Israeli Ministry of Science and Technology.